# The Sobering Reality of Perovskite/Si Tandem Solar Cells under Realistic Operating Conditions

*Moritz H. Futscher, Bruno Ehrler*

Center for Nanophotonics, AMOLF, Science Park 104, 1098 XG Amsterdam, The Netherlands

**Corresponding Author**

* ehrler@amolf.nl

Perovskite/Si tandem solar cells have the potential to considerably out-perform conventional solar cells. Under standard test conditions, perovskite/Si tandem solar cells already outperform the Si single junction. Under realistic conditions, however, as we show, those tandem solar cells are hardly more efficient than the Si cell alone. We model the performance of realistic perovskite/Si tandem solar cells under real-world climate conditions, by incorporating parasitic cell resistances, non-radiative recombination, and optical losses into the detailed-balance limit. We show quantitatively that optimizing these parameters in the perovskite top cell, perovskite/Si tandem solar cells reach an efficiency advantage of up to 14% absolute, even while leaving the Si cell untouched. Despite the rapid efficiency increase of perovskite solar cells, our results emphasize the need for further material development, careful device design, and light management strategies, all necessary for highly efficient perovskite/Si tandem solar cells.

## Introduction

Owing to the rapid increase in power conversion efficiency, metal-halide perovskite solar cells have become an auspicious candidate for cost-efficient tandem solar cells in combination with highly-optimized Si solar cells.[1–5] In a tandem configuration, a perovskite cell is stacked on top of a Si cell to absorb the high-energy part of the solar spectrum whereas the transmitted light is absorbed in the Si bottom cell. In doing so, the theoretical Shockley-Queisser limit, based on detailed balance, can be increased from 34% for a single-junction solar cell to 45% for a tandem solar cell from two subcells.[6–9]

Numerous perovskite/Si tandem solar cells have been reported in series-connected, four-terminal, and module tandem configurations, increasing the efficiency of the Si subcell alone.[10–18] With a record efficiency of 26.4%,[19] perovskite/Si tandem solar cells almost match the current record efficiency of Si solar cells of 26.6%.[20] Yet, even the best perovskite/Si tandem solar cells show only around half the efficiency of the detailed-balance efficiency limit. The efficiency is reduced due to parasitic absorption, non-radiative recombination ($J_{NR}$), undesirable series resistance ($R_S$) and shunt resistance ($R_{SH}$), and optical losses. Furthermore, we recently showed that the power conversion efficiency of perovskite/Si tandem solar cells is strongly affected by the considered tandem configuration, spectral variations, and temperature changes.[21] These considerations show that the interplay of real-world climate conditions, and realistic solar cell parameters has to be considered for an approximation of the performance of these cells. So far, however, understanding of the behaviour of realistic perovskite/Si tandem solar cells under realistic conditions is lacking.

Here we develop a model to simulate state-of-the-art perovskite solar cells by integrating non-radiative recombination, parasitic resistances, and optical losses into the detailed-balance model. Together with Si bottom cells we simulate realistic perovskite/Si tandem solar cells under real-world climate conditions to predict the potential power yield. We show that the power conversion efficiency of perovskite/Si tandem solar cells is hardly better than that of the single-junction Si solar cell under realistic conditions, even when using the best available perovskite and silicon solar cells to date (efficiency advantage < 1.5% absolute). We also find that the three tandem connection schemes (connected electrically in series, on the module level, or as independent four-terminal devices) show almost identical efficiency values (< 1.5% absolute difference). Only when reducing the parasitic absorption in the contacts, the tandem solar cells start to out-perform the single-junction Si cell by 1.7-3.2% absolute. Finally, we show how a reduction in non-radiative recombination, optimized series and shunt resistance, and a reduction in optical losses for the perovskite cell could boost the efficiency advantage of the tandem cell up to 14.1% absolute (13.0% absolute for series tandem) compared to the single-junction cell. For these optimized cells, the connection scheme, as well as the climate conditions, become more important compared to non-optimized cells.

**Results and discussion**

Including realistic solar-cell parameters such as non-radiative recombination and parasitic resistances into the detailed-balance limit calculations allows for simulating realistic solar cell performance under real-world climate conditions (temperature, irradiance, and spectrum). In the detailed-balance limit, all recombination is radiative, all light above the bandgap is absorbed, and there are no optical losses (external quantum efficiency (EQE) = 1). Knowledge of the bandgap, temperature, incoming spectrum, and intensity then allows calculating the limiting efficiency of a solar cell. We extend the detailed-balance calculations to include non-radiative recombination, series resistance, shunt resistance, and the fact that not all light above the bandgap leads to photocurrent (optical losses, EQE < 1). We model perovskite and Si solar cells based on current record efficiency devices (≥ 1 cm²).[22,23] To extract the parameters of these realistic (record) cells we use our modified detailed-balance limit to fit the current-voltage characteristics[24] (see Supplementary Information (SI) S1 for a full description of the model). To simulate real-world climate conditions we use solar spectra, irradiance, and temperatures measured in Utrecht, The Netherlands[25] and in Denver, Colorado, US[26] in 2015 at an interval of 30 min during daylight hours.

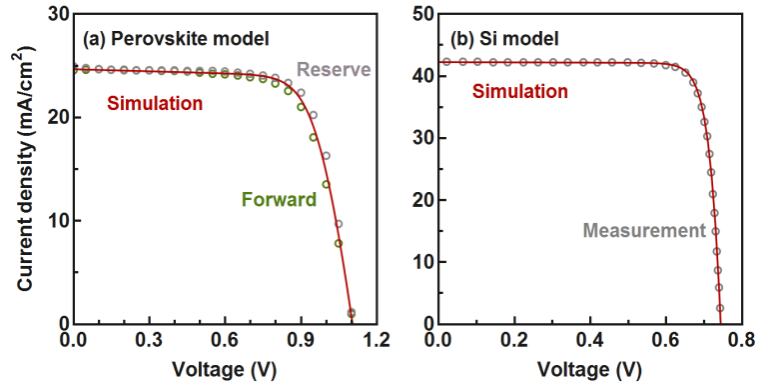

**Figure 1.** Modelled current-voltage characteristics of record efficiency (a) perovskite and (b) Si solar cells. The circles correspond to the measured data of (a) the record efficiency perovskite solar cell[22] and (b) a high-efficient Si solar cell[23]. The fit parameters are summarized in Table 1.

We fit our model to the current-voltage characteristics of record-efficiency perovskite and Si solar cells as shown in Figure 1. We include different mechanisms for non-radiative recombination for the Si and perovskite subcells. To model the Si cell we take Auger[27] recombination and a non-radiative diffusion current of minority carriers into account.

Since most of the perovskite layer is depleted,[28–30] we assume the dominating recombination mechanism to be recombination from the space charge region. As a result, the dark current of the perovskite and the Si solar cell have different dependences on temperature, irradiance, and applied voltage (see SI S2 and S3 for details). The fitted parasitic resistances and dark current densities are summarized in Table 1. Optical losses such as reflection and parasitic absorption are included by fitting the EQE of the record Si and perovskite subcells. To account for the transparent contact of the perovskite top cell, we (optimistically) assume that it absorbs 10% of the incoming light prior to reaching the silicon subcell, with additional absorption in the blue-UV region of the spectrum (see SI S4).

| | $R_S$ ($\Omega$ cm$^2$) | $R_{SH}$ ($\Omega$ cm$^2$) | $J_{NR}$ (pA/cm$^2$) | $J_{SC}$ (mA/cm$^2$) | $V_{OC}$ (V) | FF (%) | $\eta$ (%) |
|---|---|---|---|---|---|---|---|
| Perovskite | 3.10 | 1500 | 28.50 | 24.67 | 1.104 | 72.3 | 19.7 |
| Si | 0.15 | 5000 | 0.01 | 42.25 | 0.744 | 84.6 | 26.6 |

**Table 1.** Fitted solar cell parameters and performance of modelled perovskite and Si solar cells.

Using these modelled perovskite and Si subcells we calculate the efficiency for current-matched series, voltage-matched module, and unconstrained four-terminal tandem assembly strategies following previous work[21] (see SI S4 for details). For the series tandem the optical thickness of the perovskite top cell was set to 80% in order to current-match the perovskite top cell and the Si bottom cell. For the module tandem the ratio between the number of perovskite top cells and Si bottom cells was set to 1/1.38 in order voltage-match one perovskite top cell with 1.38 Si bottom cells. We calculate the efficiencies of single-junction cells and different tandem configurations under standard test conditions (AM1.5G, 1 kW/m$^2$, 25 °C) and also under real-world climate conditions averaged over an entire year (see Table 2). The average efficiency is weighted with the incoming intensity, to allow for a calculation of the integrated power conversion efficiency over the year.

| | Si cell | Series | Module | Four–terminal |
|---|---|---|---|---|
| STC | 26.6 % | 27.3 % | 28.4 % | 28.4 % |
| NL | 24.4 % | 23.9 % | 25.4 % | 25.4 % |
| CO | 24.8 % | 24.9 % | 26.2 % | 26.3 % |

**Table 2**. Intensity-weighted power conversion efficiency over an entire year for the three perovskite/Si tandem configurations and the Si solar cell under standard test conditions (STC, AM1.5G, 1 kW/m$^2$, 25 °C) and under real-world conditions at two locations with distinctively different climate conditions: Utrecht, The Netherlands (NL)[25] and Denver, Colorado (CO)[26].

Rather surprisingly, combining record efficiency perovskite and Si solar cells in a tandem configuration would increase the efficiency only marginally compared to the single-junction Si solar cell alone. In fact, in the series tandem configuration, the efficiency is reduced by 0.5% absolute in the Netherlands, due to the stronger dependence on the incoming spectrum and irradiance. For this tandem configuration, the additional power generated in the perovskite top cell does not outweigh the losses introduced by placing the perovskite cell on top of the Si cell. We note that the difference between the different tandem configurations is less than 1.5% absolute, lower than expected from ideal cells. The losses compared to ideal cells are caused by shading part of the light from the silicon cell in the perovskite solar cell contacts, but also from electrical losses as the electrical characteristics of the perovskite cell are not as highly optimized as those of the Si cell. These effects counteract some of the tandem specific losses due to spectrum and temperature changes rendering the different tandem configurations similar in efficiency.

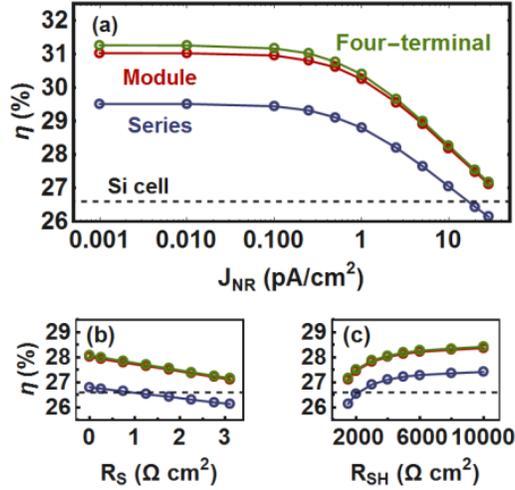

**Figure 2**. Effect of optimizing (a) non-radiative recombination ($J_{NR}$), (b) series resistance ($R_S$), and (c) shunt resistance ($R_{SH}$), of the perovskite top cell on the intensity-weighted power conversion efficiency over a year for the three perovskite/Si tandem configuration calculated using solar spectra and temperatures measured in Utrecht, The Netherlands[25]. The dashed line indicates the performance of the Si bottom cell at standard test conditions. The calculations assume no parasitic absorption in the perovskite cell contacts.

The efficiency of perovskite solar cells and perovskite/Si tandem solar cells as a function of solar irradiance, for ideal cells (Figure 3 (a)) and with realistic solar cell parameters such as $J_{NR}$, $R_S$ and $R_{SH}$ (Figure 3 (b)) shows that the difference in efficiency between ideal and realistic cells is most prominent in the low-intensity region. Crucially, electrical losses are particularly harmful at low light intensity and are hence typically underestimated when the solar cells are evaluated under standard test conditions. The strong decrease in efficiency at low irradiance is due to unfavourable $R_{SH}$ whereas the effects of $R_S$ and $J_{NR}$ are more prominent at high irradiance (see SI S5). In the following we will examine how optimizing the parameters of the perovskite cell, such as parasitic absorption in the contacts, improving the electrical characteristics, and eliminating all optical losses, can lead to a massive increase in efficiency compared to both the single-junction Si cell, and the tandem cells made from state-of-the-art subcells.

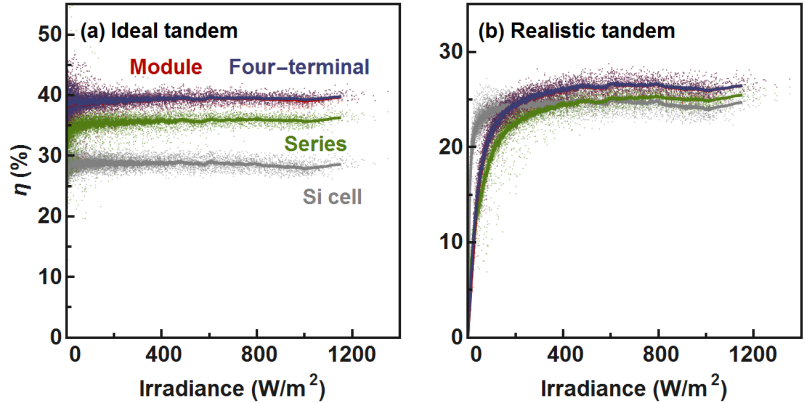

**Figure 3.** Efficiency of the three perovskite/Si tandem configurations and the perovskite solar cell under real-world conditions as a function of irradiance calculated using solar spectra and temperatures measured in Utrecht, The Netherlands[25] for (a) ideal and (b) record efficiency Si and perovskite subcells. The solid line represents a moving average of the data.

Assuming no parasitic absorption in the transparent contact of the perovskite cell leads to an increase in the efficiency of tandem solar cells between 1.7 and 3.2% absolute compared to the Si cell alone, depending on the considered tandem configuration and location. This increase stems from the increased current in the Si subcell. While this scenario might be unrealistic for typical transparent contacts, novel contacting schemes as well as nanowire contacts have been proposed to reduce the overall optical losses in the transparent contact.[31–35] For simplicity we assume no parasitic contact absorption in the perovskite top cell in the following discussion.

To understand the most practical ways to increase the efficiency of perovskite/Si TSCs we systematically change the parameters $J_{NR}$, $R_S$, and $R_{SH}$ of the perovskite top cell in our model (Figure 2). For each calculation the tandem thickness of the perovskite top cell for the series tandem, and the ratio of Si to perovskite cells for the module tandem were optimized for maximum efficiency. Reducing the non-radiative recombination to 1 fA/cm$^2$ of the perovskite subcell leads to a strong increase in power-conversion efficiency of up to 4.4% absolute. Optimizing parasitic resistances can further increase the efficiency by up to 0.9% absolute for $R_S = 0$ $\Omega$ cm$^2$ and up to 1.3% absolute for $R_{SH} = 10{,}000$ $\Omega$ cm$^2$.

These changes in efficiency are different for the different tandem configurations. Due to the  non-ideal perovskite bandgap of 1.49 eV for the formamidinium lead iodide perovskite,[22] the series tandem gains the least in efficiency by optimizing the perovskite subcell compared to the module and the four-terminal tandem. This is especially evident when optimizing $J_{NR}$ and $R_S$ of the perovskite top cell where the module and the four-terminal tandem gain about 0.5% absolute more in efficiency than the series tandem. The reduction of non-radiative recombination is slightly more beneficial for the four-terminal tandem than for the module tandem since the module tandem is subject to voltage-matching. However, the module tandem benefits notably more from a decrease in $R_S$ since a change in $R_S$ strongly changes the slope of the current-voltage characteristic close to the open-circuit voltage.

So far we used the EQE of the record cells to account for optical losses due to reflection, parasitic absorption and potential electrical losses. Assuming an ideal EQE of 1 for the perovskite subcell the efficiency of the tandem cells can be further increased by about 0.3% for the series tandem and about 2.6% for the module and the four-terminal tandem. The series tandem is much less affected since the gain in current in the perovskite cell here is balanced by less light that it transmitted to the silicon cell. For the module and four-terminal tandem cell, the perovskite cell was optically thick to start with, so the additional EQE gain is almost exclusively a current gain for the overall tandem cell.

After optimizing the perovskite subcell there is a strong difference in efficiency between the series tandem, and the module and four-terminal tandem using a perovskite top cell with a bandgap of 1.49 eV, due to the need for a semi-transparent perovskite layer. This difference in efficiency between the tandem configurations almost vanishes by using a perovskite cell with an ideal bandgap of 1.73 eV.

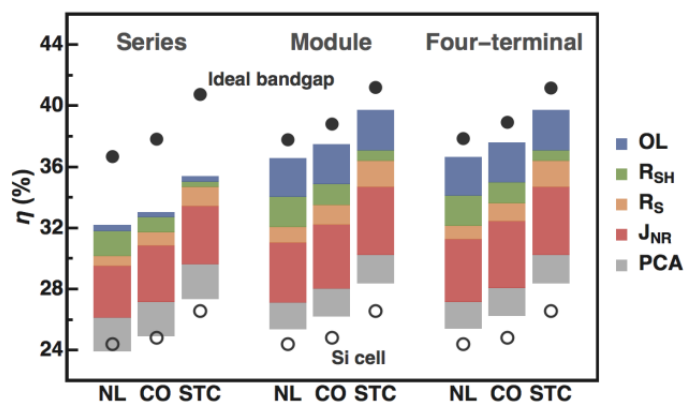

**Figure 4.** Effects of parasitic contact absorption losses (PCA), non-radiative recombination ($J_{NR}$), series resistance ($R_S$), shunt resistance ($R_{SH}$), and optical losses (OL) on the intensity-weighted power conversion efficiency over a year for the three perovskite/Si tandem configurations, calculated using standard test conditions (STC) and solar spectra and temperatures measured in Utrecht, The Netherlands (NL)[25] and in Denver, Colorado (CO)[26]. The empty circles indicate the efficiency of the Si bottom cell alone. The filled circles indicate the tandem efficiency when using an optimized perovskite top cell with an ideal bandgap of 1.73 eV.

Taken together, optimizing the perovskite cells can lead to an efficiency advantage of a tandem cells up to 13.0% absolute for the series tandem and 14.1% absolute for the four-terminal tandem, even when leaving the silicon cell untouched (Figure 4). When optimized, the perovskite solar cell has an efficiency of 32.1% for a bandgap of 1.49 eV, and 28.4% for the ideal bandgap under standard operating conditions (see Figure S15 in the SI for details).

The efficiencies for the tandem configurations are different for the different locations. The Netherlands comprises a larger fraction of low-irradiance solar spectra throughout the year compared to Colorado.[21] This difference in irradiance leads to a higher annual-averaged power conversion in Colorado than in the Netherlands (Figure 3). For the same reason, the reduction of $J_{NR}$ and the improvement of $R_S$ have a larger effect in Colorado whereas the increase $R_{SH}$ has a larger effect in the Netherlands.

**Conclusions**

We simulated realistic perovskite/Si tandem solar cells under real-world climate conditions. Our results show that even with the current record efficiency perovskite and Si solar cells, a tandem cell would only marginally improve the efficiency of the Si cell alone under realistic operating conditions. For the series tandem cell, the efficiency of the tandem configuration can even be lower than the single-junction Si cell. When all parasitic absorption in the transparent contacts of the perovskite cell is reduced, the advantage in efficiency for the tandem cells accounts to 1.7 – 3.2% absolute compared to the Si single–junction cell. This is far less than expected for ideal cells under ideal conditions, as the realistic cells are also more sensitive to the local climate conditions, in particular the average irradiance. We show that further optimizing parasitic cell resistances, non-radiative recombination, and optical losses of the perovskite top cell can boost the efficiency advantage of the tandem cells by up to 14.1% absolute, surpassing the single-junction Shockley-Queisser limit for all three tandem configurations under realistic operation conditions. We show that reducing the non-radiative recombination, optimizing electrical non-idealities, and optical losses all have a large potential for efficiency gain. Optimizing the perovskite cell also leads to an increased efficiency difference between the series-connected tandem cells and the four-terminal and module tandem cells.

Perovskite cells have shown a breath-taking development in efficiency over recent years. Yet, our results highlight the need for a concerted effort in material development (reducing non-radiative recombination and tuning the bandgap), device design (optimizing cell resistances), and light management strategies (reducing optical losses and developing transparent contacts) to further increase the efficiency of perovskite cells, and develop highly efficient perovskite/Si tandem solar cells.

**Acknowledgements**

The authors thank W.G.J.H.M. van Sark and A. Louwen from the Utrecht Photovoltaic Outdoor Test facility (UPOT), and the NREL Solar Radiation Research Laboratory for providing spectral irradiance and temperature data and Erik Garnett for carefully reading and commenting on the manuscript. This work is part of the research programme of the Netherlands Organisation for Scientific Research (NWO).

# SUPPLEMENTARY INFORMATION FOR

# The Sobering Reality of Perovskite/Si Tandem Solar

# Cells under Realistic Operating Conditions

*Moritz H. Futscher, Bruno Ehrler\**

Center for Nanophotonics, AMOLF, Science Park 104, 1098 XG Amsterdam, The Netherlands

AUTHOR INFORMATION

**Corresponding Author**

\* ehrler@amolf.nl

## S1 REALISTIC SOLAR CELL MODEL

To model realistic solar cells, we incorporate non-radiative recombination into the detailed-balance limit[1] by analytically solving the current-voltage characteristics for a *p-n* junction solar cell consisting of a quasi-neutral *p*-type region, a space charge region, and a quasi-neutral *n*-type region using the depletion and the superposition approximation following Jenny Nelson[2]. The depletion approximation assumes that the electric field is zero in the quasi-neutral regions and varies linearly across the depletion region, the junction contains no free carriers, and the quasi-Fermi levels are constant across the depleted region. This allows to decouple solutions for the quasi-neutral regions and the space charge region. The superposition approximation furthermore allows to decouple the effect of bias from the effect of illumination by assuming that the recombination rates in the quasi-neutral regions are linear in the minority-carrier density. In addition, we assume that the recombination rate is constant within the depletion region and Shockley-Read-Hall recombination through a single trap state is the dominant recombination process. This results in two non-radiative recombination currents originating from the quasi-neutral layer and the space charge region of the *p-n* junction with different ideality factors.

$$J_{Non-Rad} = J_{NR,1}\left(e^{\frac{qV}{k_BT}} - 1\right) + J_{NR,2}\left(e^{\frac{qV}{2k_BT}} - 1\right) \tag{1}$$

with

$$J_{NR,1} = q\,n_i{}^2\left(\frac{D_n}{N_a\,L_n} + \frac{D_p}{N_d\,L_p}\right)$$

and

$$J_{NR,2} = \frac{q\,n_i\,w_{SCR}}{\sqrt{\tau_n\,\tau_p}}.$$

The first term of $J_{Non-Rad}$ is the recombination current density from the quasi-neutral regions due to minority-carrier diffusion and the second term of $J_{Non-Rad}$ the recombination current density from the space charge region. Here $q$ is the elementary charge, $n_i$ the intrinsic charge carrier density, $D_n$ and $D_p$ the diffusion constants for electrons and holes, $N_a$ and $N_d$ the doping densities of acceptor and donor, $L_n$ and $L_p$ the diffusion lengths for electrons and holes, $V$ the applied voltage, $k_B$ the Boltzmann constant, $T$ the temperature of the cell, $w_{SCR}$ the width of the space charge region, and $\tau_n$ and $\tau_p$ the lifetimes of electrons and holes. The width of the space charge region is given by

$$w_{SCR} = \sqrt{\frac{2\,\epsilon}{q}\left(\frac{1}{N_a} + \frac{1}{N_d}\right)(V_{bi} - V)} \tag{2}$$

where $\epsilon$ is the permittivity and $V_{bi}$ the built-in bias. Equation 2 is only defined for $V_{bi} > V$. Auger recombination is taken into account as[3]

$$J_{Auger} = J_A \left(e^{\frac{3\,q\,V}{2\,k_B T}} - 1\right)$$

with

$$J_A = q\,L\,C\,n_i^3$$

where $L$ is the thickness of the material and C the Auger coefficient. The non-radiative recombination current densities are now added to the detailed-balance limit as

$$J = J_G - J_{Rad} - J_{Non-Rad} - J_{Auger}$$

where $J$ is the total current density generated by the solar cell, $J_G$ the generated photocurrent density, and $J_{Rad}$ the radiative recombination current density. The radiative recombination is given by

$$J_{Rad} = f_w \frac{2\pi q}{c^2 h^3} \int_{E_G}^{\infty} \frac{E^2}{e^{\frac{E - qV}{k_B T}} - 1}\, dE$$

where $c$ is the speed of light, $h$ Planck's constant, $f_w$ a geometrical factor taking into account the angle upon which the radiative recombination emits from the cell, and $E_G$ the bandgap. In case of $E_G - qV \gg kT$, the radiative recombination can be approximated as[4]

$$J_{Rad} = J_R \left( e^{\frac{q\,V}{k_B T}} - 1 \right)$$

with

$$J_R = f_w \frac{2\pi q}{c^2 h^3} \int_{E_G}^{E_{max}} \frac{E^2}{e^{\frac{E}{k_B T}} - 1} \, dE,$$

which retrieves the ideal diode equation. For single-junction solar cells we set $f_w$ to 1 meaning that the cell only emits via its front side. To account for additional optical losses such as parasitic absorption and reflection, the external quantum efficiency (EQE) is taken into account when calculating the photo-generated current density as

$$J_G = q \int_{E_{min}}^{E_{max}} \Gamma(E) \, \text{EQE}(E) \, dE$$

where $\Gamma$ is the photon flux of the incident solar spectrum. Internal cell resistances due to parasitic series and shunt resistance reducing the fill factor of the solar cell are included as

$$J = J_G - J_R \left( e^{\frac{q\,(V + J\,R_S)}{k_B T}} - 1 \right) - J_{NR,1} \left( e^{\frac{q\,(V + J\,R_S)}{k_B T}} - 1 \right) -$$

$$J_{NR,2} \left( e^{\frac{q\,(V + J\,R_S)}{2\,k_B T}} - 1 \right) - J_A \left( e^{\frac{3\,q\,(V + J\,R_S)}{2\,k_B T}} - 1 \right) - \frac{V + J\,R_S}{R_{sh}} \qquad (3)$$

where $R_s$ is the series resistance and $R_{sh}$ the shunt resistance. The current-voltage characteristic was calculated by numerically solving Equation 3 for $J(V)$. Temperature and irradiance effects are included in the model by adding material specific parameters for perovskite and Si as discussed in S2 and S3.

## S2 SI SOLAR CELL MODEL

Since no current-voltage characteristics of the current record Si solar cell with an efficiency of 26.6%[5] has been published to date, we fit the current-voltage characteristic of the Si solar cell with an efficiency of 26.3%[6] and then change internal cell resistances to model the record Si solar cell with an efficiency of 26.6%. The EQE of the Si solar cell with an efficiency of 26.3% is used to account for optical losses and parasitic absorption (see Figure S1). To account for spectral regions outside the published range we set the regions below 300 and above 1200 nm to 0% EQE.

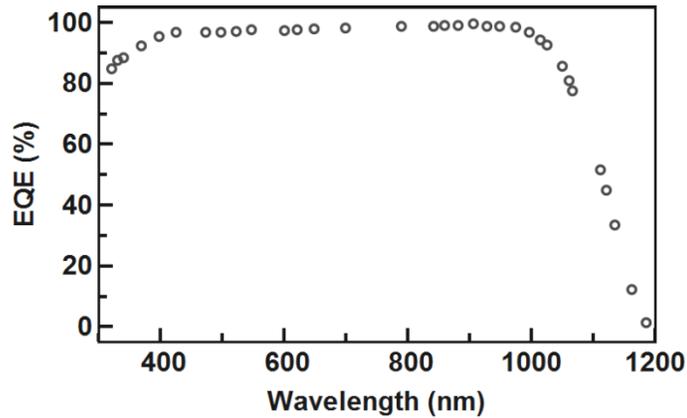

**Figure S1.** External quantum efficiency of a Si solar cell[6] with an area of 180 cm² and an efficiency of 26.3% used to account for optical losses.

Since Si has an indirect bandgap the diffusion length of electrons and holes are long compared to the width of the space charge region. Consequently, there is only very little recombination originating from the space charge region.[2] Non-radiative recombination current density of the Si solar cell is thus calculated by approximating Equation 1 as

$$J_{Non-Rad} = J_{NR,1}\left(e^{\frac{qV}{k_BT}} - 1\right)$$

where $J_{NR,1}$ is calculated as

$$J_{NR,1} = \alpha\, n_i{}^2$$

with the fitting parameter $\alpha$ to adjust the amount of non-radiative recombination due to diffusion limited current by minority carriers as well as surface and contact recombination.[7] The temperature dependence of the intrinsic charge carrier density is implemented in our model as[8]

$$n_i = 5.29\ 10^{25} \left(\frac{T}{300\ K}\right)^{2.54} e^{-\frac{6726\ K}{T}}\ m^{-3}$$

which is shown in Figure S2a. The temperature dependence of the bandgap is calculated using Varshni's empirical equation as[9]

$$E_G = E_{G,0\ K} - \frac{\alpha\ T^2}{T + \beta}$$

with $E_{G,0\ K}$, $\alpha$, and $\beta$ set to 1.17 eV, 4.73 $10^{-4}$ K$^{-1}$, and 636 K, respectively.[10] The bandgap of the simulated Si solar cell as a function of temperature can be seen in Figure S2b. The thickness of the Si wafer[6] is set to L = 165 μm and the temperature dependence of the Auger coefficient shown in Figure S2c is calculated as[11]

$$C = C_n + C_p = 3.79\ 10^{-43} \sqrt{\frac{T}{300\ K}}\ \frac{m^6}{s}$$

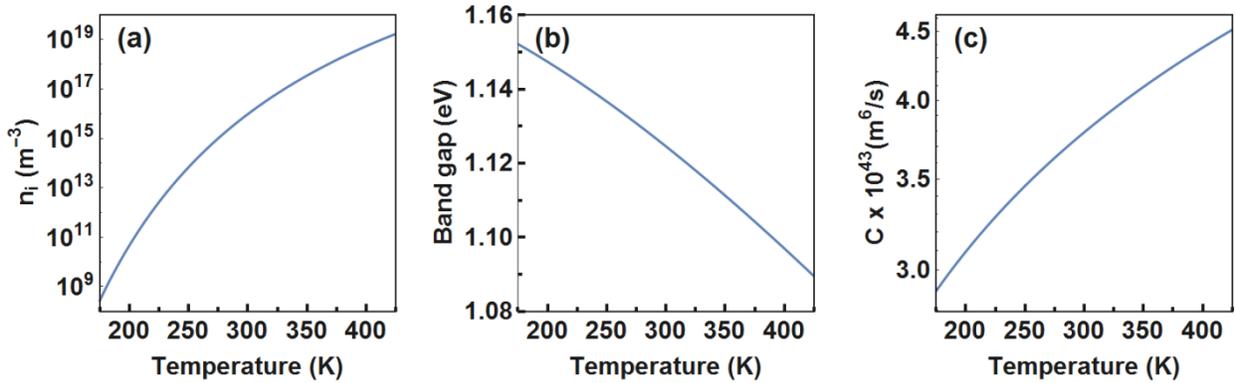

**Figure S2.** Temperature dependence of (a) the intrinsic charge carrier concentration, (b) the bandgap, and (c) the Auger coefficient used to model the record Si solar cell.

The current-voltage characteristic of the modelled record Si cell can thus be written as

$$J = J_G - J_R \left( e^{\frac{q\,(V+J\,R_S)}{k_B T}} - 1 \right) - \alpha\, n_i{}^2 \left( e^{\frac{q\,(V+J\,R_S)}{k_B T}} - 1 \right) - J_A \left( e^{\frac{3\,q\,(V+J\,R_S)}{2\,k_B T}} - 1 \right) - \frac{V+J\,R_s}{R_{sh}}$$

which only has three fitting parameters: $\alpha$, $R_s$ and $R_{sh}$. The fitting parameters used to model the record Si solar cell are shown in Table 1 of the main text, which agree with values for high-efficient silicon solar cells.[12]

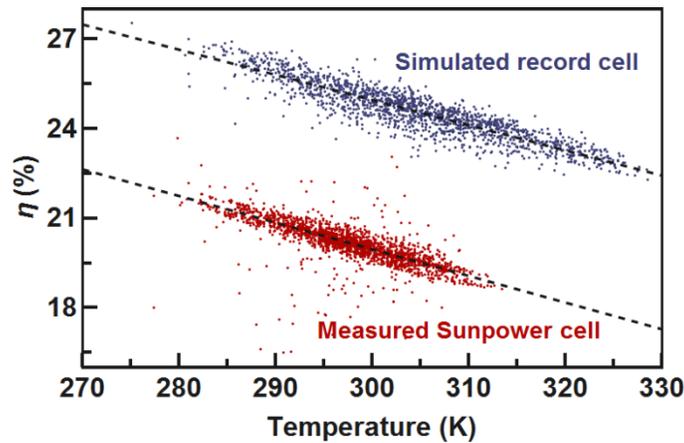

**Figure S3.** Performance of a Sunpower Si cell measured at the AMOLF outdoor test facility[13] together with the simulated performance of the record Si single-junction solar cell at real-world climate conditions as a function of temperature. Only data with a solar irradiance higher than 500 W/m² are shown. The dotted lines correspond to linear regressions of the data used to quantify the temperature dependence.

In order to test the accuracy of the temperature dependence of our model we compare simulated record Si single-junction efficiencies using spectral and temperature data measured in Utrecht, The Netherlands[14] in 2015 with measured efficiencies of an integrated-back contact Sunpower X-Series Si solar panel with a nominal efficiency of 21.5% measured at AMOLF in Amsterdam, The Netherlands between January 1ˢᵗ and April 30ᵗʰ. The simulated Si cell has a temperature dependence of -0.84% absolute per 10 K whereas the measured one has a temperature dependence of -0.89% absolute per 10 K. The temperature-dependent efficiency of the simulated and the measured Si solar cell is shown in Figure S3 for solar spectra with an irradiance higher than 500 W/m$^2$ showing that the simulation fits the temperature dependence of a measured Si solar cell under real-world conditions reasonably well.

## S3 PEROVSKITE SOLAR CELL MODEL

To model the record perovskite solar cell we fit the current-voltage characteristic of the record perovskite solar cell with an area of 1 cm$^2$ and an efficiency of 19.7%.[15] Even though efficiencies above 22% have been published,[16] these were on smaller cells, and with non-stabilized efficiency parameters. The EQE of the same perovskite solar cell is used to account for losses due to optical losses and parasitic absorption (see Figure S4). We distinguish between transmitted light (re-absorbed by the Si cell in a tandem configuration) and parasitic absorption and reflection as described in section S4. To account for spectral regions outside the published range we set the regions below 300 and above 900 nm to 0% EQE.

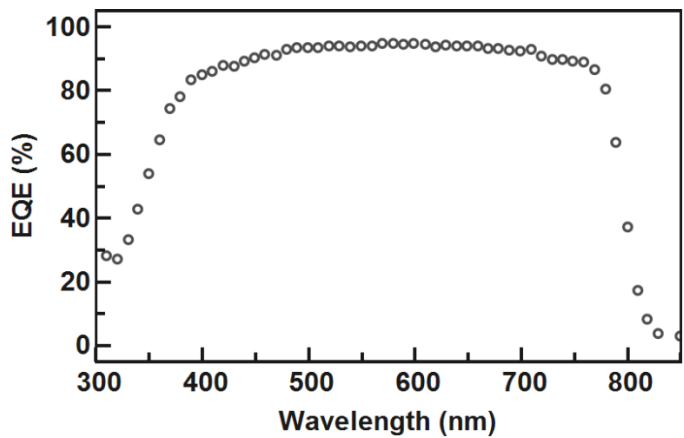

**Figure S4.** External quantum efficiency of the record perovskite solar cell[15] with an area of 1 cm$^2$ and an efficiency of 19.7% used to account for optical losses.

Since perovskite strongly absorbs light, the perovskite film thickness in solar cells is typically on the order of a few hundred nanometre. Thus, most of the perovskite layer is depleted.[17,18] Indeed it was found that the perovskite layer is fully depleted at short-circuit conditions and at negative (reverse) bias.[19] The depletion layer was found to decrease when applying a positive (forward) bias.[20] As discussed in S1, recombination from the space charge region has an ideality factor of 2, which has often been observed for perovskite solar cells.[21] If the recombination would be limited by recombination of minority carriers outside the space charge region, surface recombination, or contact recombination, the ideality factor would be close to 1. We hence assume non-radiative recombination in the perovskite solar cell to be dominated by recombination from the space charge region.[22] By approximating Equation 1, non-radiative recombination current density of the perovskite solar cell is calculated as

$$J_{Non-Rad} = J_{NR,2} \left( e^{\frac{qV}{2k_B T}} - 1 \right)$$

where $J_{NR,2}$ is calculated as

$$J_{NR,2} = \beta \frac{n_i}{\tau} \sqrt{(V_{bi} - V)}$$

with the fitting parameter $\beta$ to adjust the amount of non-radiative recombination. Due to a lack of publications we assume that the lifetime for electrons and holes are the same. The built-in bias is proportional to the intrinsic charge carrier density and the temperature as

$$V_{bi} \propto T \ln \left( \frac{1}{n_i^2} \right)$$

since the built-in bias is typically a bit higher than the open-circuit voltage we set it to 1.3 V at 25 °C, a bit higher than the open-circuit voltage of the Shockley-Queisser limit for a bandgap of 1.49 eV (see Figure S5a). The density of states in the conduction band and in the valance band are assumed to be $N_C = N_V = 3.97 \, 10^{18} \, cm^{-3}$ at 25 °C assumed by DFT calculations.[23,24] The temperature-dependent intrinsic charge carrier density is thus calculated as

$$n_i = 7.36 \; 10^{20} \; T^{3/2} \; e^{-\frac{E_G}{2 \, k_B T}} \; m^{-3}$$

which is shown in Figure S5b. The bandgap is assumed to increase by 0.35 meV per K and is calculated as[25]

$$E_G = 1.38565 \; eV + 0.00035 \frac{eV}{K} * T.$$

The bandgap of the simulated Si solar cell as a function of temperature can be seen in Figure S5c.

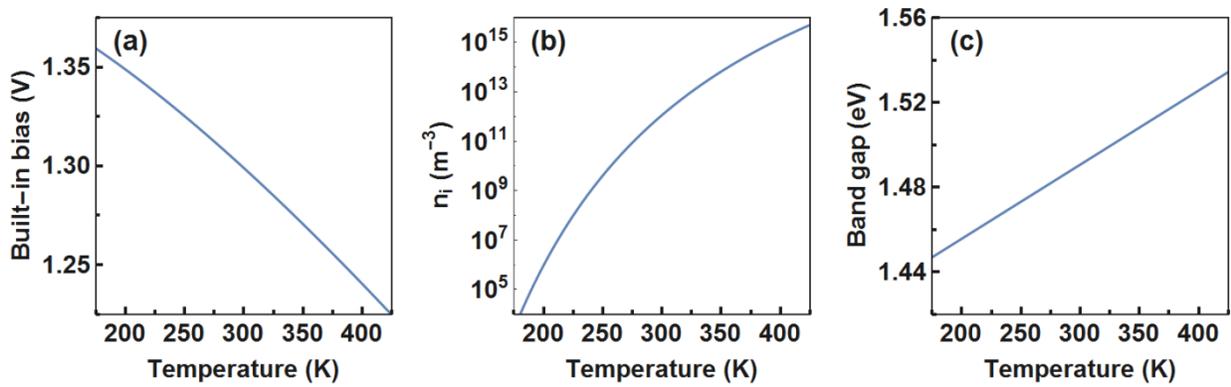

**Figure S5.** Temperature dependence of (a) the built-in bias, (b) the intrinsic charge carrier concentration, and (c) the bandgap used to model the record perovskite solar cell.

Due to the low intrinsic charge carrier density in perovskites, the photo-generated charge carrier density has to be taken into account. The effect of irradiance on non-radiative recombination is included by using a charge carrier lifetime that depends on the solar irradiance $G$. The irradiance dependent charge carrier lifetime was included in the model by fitting current-voltage characteristics of methylammonium lead iodide perovskites measured at different light intensities.[26] The current-voltage characteristics were first fitted at 1 Sun illumination to retrieve values for the parasitic absorption, resistance, and non-radiative recombination. At 1 Sun, the charge carrier lifetime was set to 1 μs, in close agreement with many reports for this material.[27,28] These values were then used to fit current-voltage characteristics at different irradiances by changing only the irradiance and the charge carrier lifetime. The fitted current-voltage characteristics together with the obtained charge carrier lifetimes are shown in Figure S6. The irradiance dependent charge carrier lifetime to model the record perovskite cell was then calculated as

$$\tau = 2.63 \left(\frac{G}{Wm^2}\right)^{-0.14} \mu s,$$

where G is the irradiance. A higher charge carrier lifetime at low irradiances would slightly decrease the effect of real-world climate conditions since it would reduce the non-radiative recombination. To test for the sensitivity of the fit on the lifetime dependence, we run our calculations for a much stronger irradiance dependence of the lifetime ($\tau = 100 \left(\frac{G}{Wm^2}\right)^{-1/1.5} \mu s$). We find that the overall efficiency of the tandem cells increases only slightly, by 0.2% absolute. We note that the values for the charge carrier lifetime were extrapolated from measurements of a methylammonium lead iodide perovskite which might differ from charge carrier lifetimes of formamidinium lead iodide perovskites.

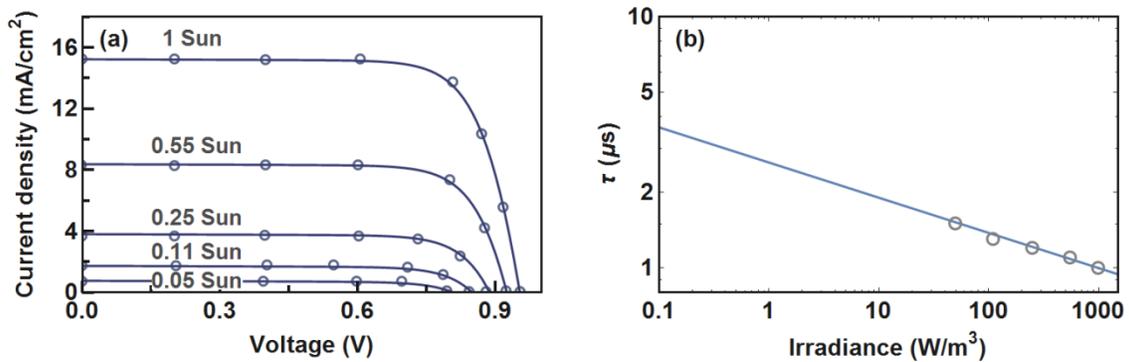

**Figure S6.** (a) Measured current-voltage characteristics[27] (a) of a perovskite solar cell at different light intensities used to extrapolate irradiance dependent charge carrier lifetimes (b) used to model the record perovskite solar cell.

Since perovskite solar cells only show insignificant Auger recombination[29] it is neglected in our model. The current-voltage characteristic of the modelled record perovskite cell can thus be written as

$$J = J_G - J_R \left( e^{\frac{q\,(V+J\,R_S)}{k_B T}} - 1 \right) - \beta\, \frac{n_i}{\tau} \sqrt{(V_{bi} - V)}$$

$$\left( e^{\frac{q\,(V+J\,R_S)}{2k_B T}} - 1 \right) - \frac{V + J\,R_s}{R_{sh}}$$

which has only three fitting parameters: $\beta$, $R_s$ and $R_{sh}$. The fitting parameters used to model the record perovskite solar cell are shown in Table 1 in the main text.

## S4 TANDEM MODEL

To calculate the efficiency of perovskite/Si tandem solar cells we apply modifications to the detailed-balance limit similar to the work of De Vos[30] and Strandberg[31]. We assume a perfect reflector on the back side of the Si bottom cell and no selective reflector between the perovskite top cell and the Si bottom cell. For simplicity we neglect reabsorption of luminescence between the two subcells. For full details of the tandem solar cell model are described in our earlier work [32].

Optical losses are included by fitting a Gaussian distribution with a full width at half maximum (FWHM) of 53 meV to the onset of the EQE spectra of the perovskite cell (Figure S7). 10% of the light with an energy below the Gaussian distribution is assumed to be uniformly absorbed by parasitic absorption in the perovskite contacts. Light above the Gaussian distribution is almost completely absorbed in the perovskite. We include a small fraction of light between the bandgap and 600 nm that is transmitted to follow published transmission curves.[33] The transmitted light by the perovskite top cell is shown in Figure S7 (b) together with the fitted Gaussian distribution.

Figure S8 shows the EQE of the perovskite and the Si subcell used to model perovskite/Si tandem solar cells with contact losses.

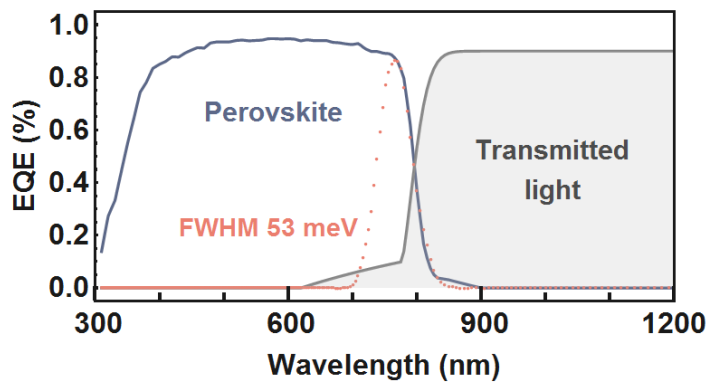

**Figure S7.** EQE model used to simulate realistic perovskite/Si tandem cells. EQE of perovskite solar cell (blue) together with the fitted Gaussian distribution (red) used to model realistic perovskite/Si tandem cells. Gray shaded areas represent the light which is transmitted to the Si bottom cell.

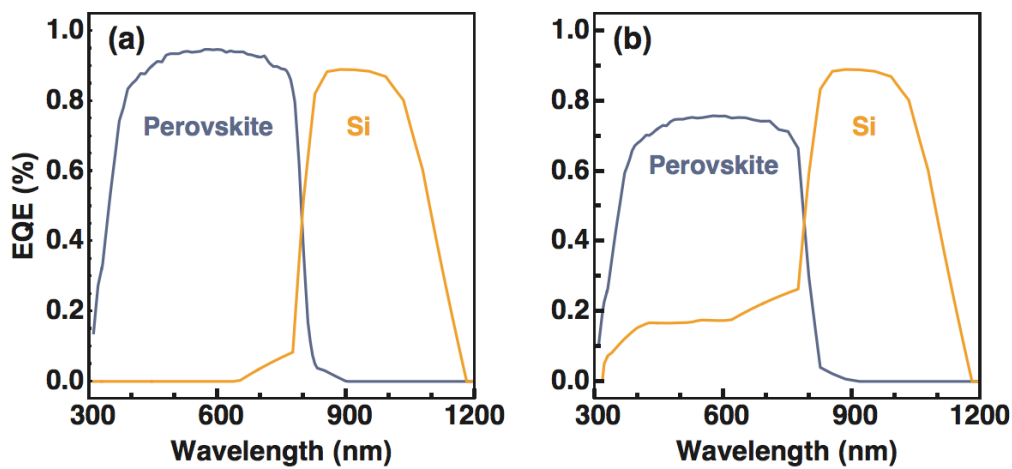

**Figure S8.** EQE of the modelled perovskite/Si tandem solar cells with record perovskite and Si subcells for (a) the module and the four-terminal tandem with an optical thick perovskite top cell and (b) for the series tandem with an optical thin perovskite top cell to achieve current matching.

When neglecting transmission losses due to parasitic absorption and optical losses from the perovskite top cell the current density in the Si bottom cell was calculated as

$$J_{G,Si} = q \int_{E_{min}}^{E_{max}} \Gamma(E) \left[ \text{EQE}_{Si}(E) - EQE_{Perovskite}(E) \right] dE$$

Where $\text{EQE}_{Si}$ is the EQE of the Si bottom cell and $\text{EQE}_{Perovskite}$ is the EQE of the perovskite top cell. For an optically thin perovskite top cell the current density was calculated as

$$J_{G,Perovskite}^{thin} = q \int_{E_{min}}^{E_{max}} \Gamma(E) \, \gamma \, EQE_{Perovskite}(E) \, dE$$

and

$$J_{G,Si}^{thin} = q \int_{E_{min}}^{E_{max}} \Gamma(E) \left[ \text{EQE}_{Si}(E) - \gamma \, EQE_{Perovskite}(E) \right] dE$$

where $\gamma$ is the absorbed light fraction of the perovskite top cell. Note that this corresponds to an ideal case where all light not converted by the perovskite top cell is transmitted to the Si bottom cell. The EQEs used to model perovskite/Si tandem solar cells without transmission losses are shown in Figure S9 to Figure S11.

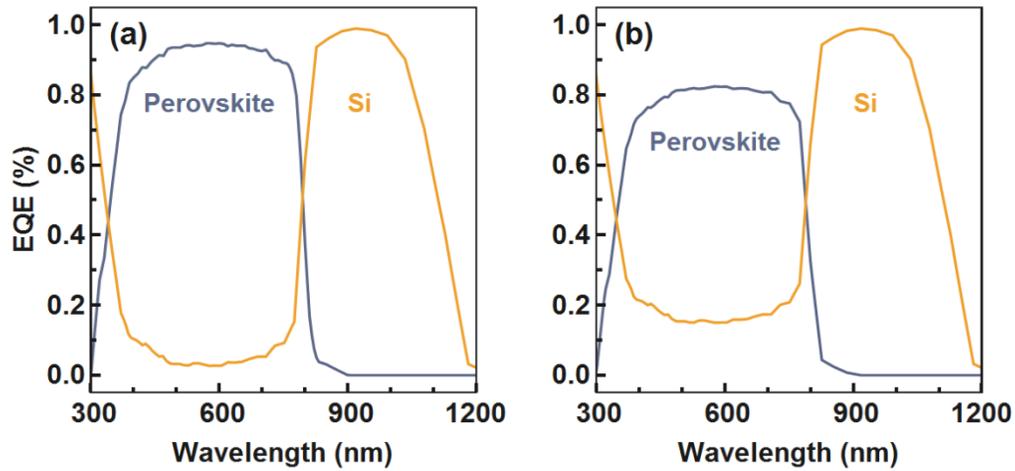

**Figure S9.** EQE of the modelled perovskite/Si tandem solar cells with record perovskite and Si subcells and no contact losses for (a) the module and the four-terminal tandem with an optical thick perovskite top cell and (b) for the series tandem with an optical thin perovskite top cell to archive current matching.

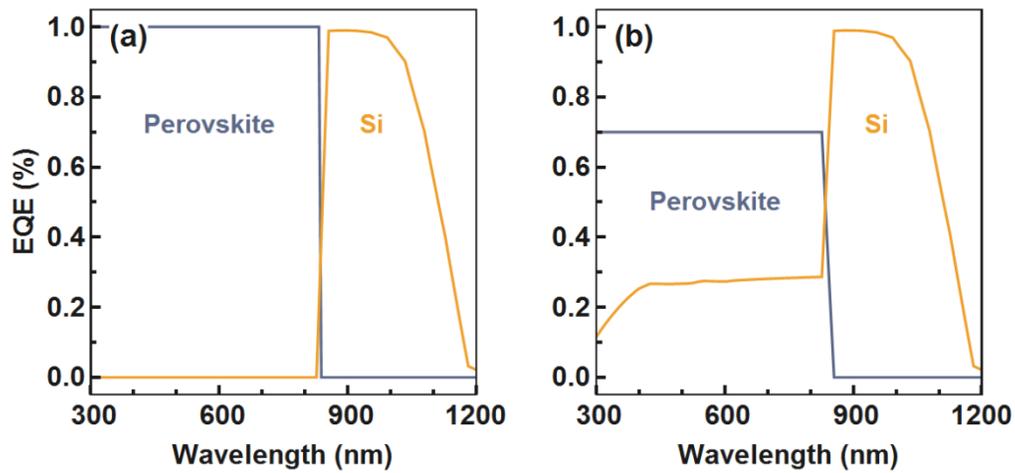

**Figure S10.** EQE of the modelled perovskite/Si tandem solar cell for the record Si subcell and a perovskite subcell with an ideal EQE of 100% and no contact losses for (a) the module and the four-terminal tandem with an optical thick perovskite top cell and (b) for the series tandem with an optical thin perovskite top cell to archive current matching.

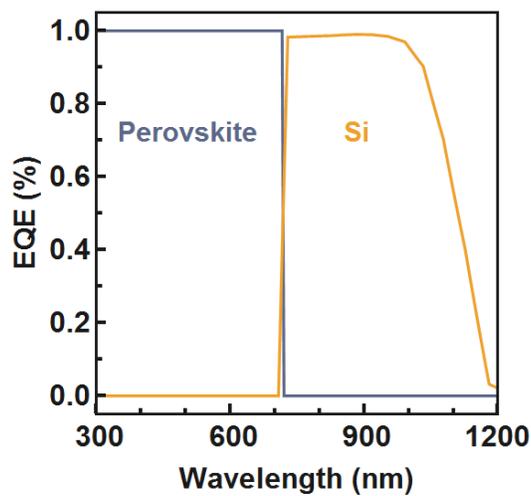

**Figure S11.** EQE of the modelled perovskite/Si tandem solar cell for the record Si subcell and a perovskite subcell with an ideal bandgap and an ideal EQE of 100%

# S7 IMPACT OF NON-RADIATIVE RECOMBINATION AND PARASITIC RESISTANCE

Figure S12 - Figure S14 show the effect of improving non-radiative recombination, series resistance and shunt resistance of the perovskite top cell on the efficiency of perovskite/Si tandem solar cells for the series, the module and the four-terminal tandem. Optimizing non-radiative recombination and series resistance results in an overall increase in efficiency most prominent at high irradiance. Optimizing shunt resistance on the other hand results in a strong increase in efficiency at low irradiance.

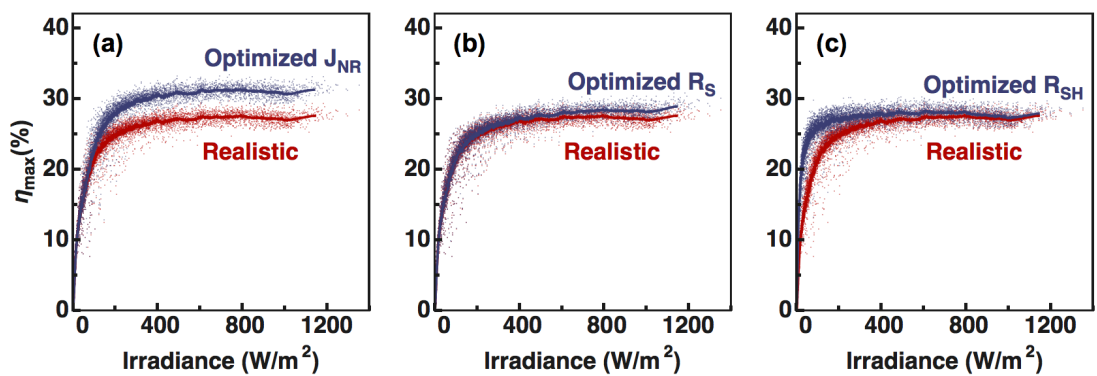

**Figure S12.** Impact of non-radiative recombination ($I_{NR}$), series resistance ($R_S$), and shunt resistance ($R_{SH}$) of the series tandem efficiency as a function of irradiance. The solid line represents a moving average of the data.

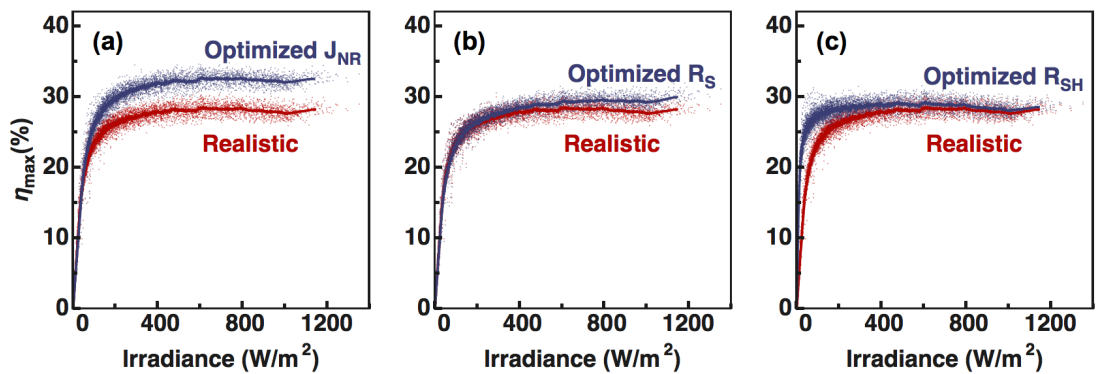

**Figure S13.** Impact of non-radiative recombination ($I_{NR}$), series resistance ($R_S$), and shunt resistance ($R_{SH}$) of the module tandem efficiency as a function of irradiance. The solid line represents a moving average of the data.

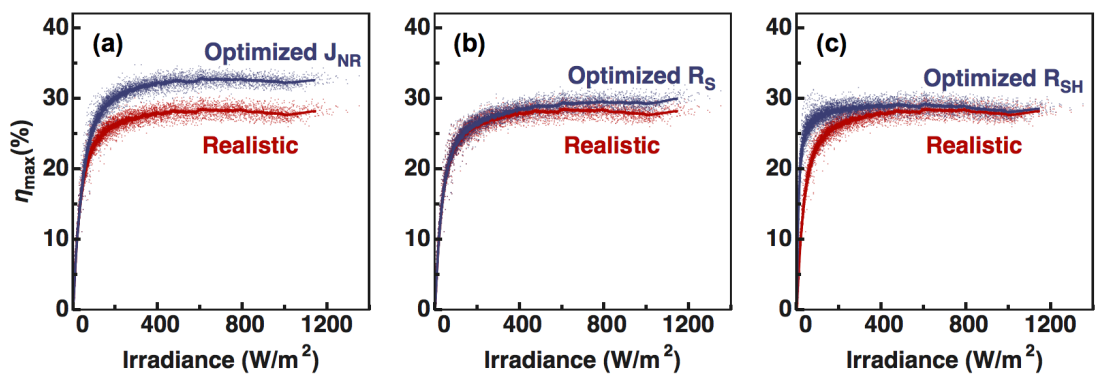

**Figure S14.** Impact of non-radiative recombination ($I_{NR}$), series resistance ($R_S$), and shunt resistance ($R_{SH}$) of the four-terminal tandem efficiency as a function of irradiance. The solid line represents a moving average of the data.

## S8 OPTIMIZING THE PEROVSKITE SOLAR CELL

Figure S15 shows the efficiency of the perovskite cell with a bandgap of 1.49 eV, when optimizing non-radiative recombination, parasitic resistances and optical losses as described in the main text. The optimized perovskite cell with a bandgap of 1.73 eV has a calculated efficiency of 28.4% using standard test conditions, 26.2% using solar spectra and temperatures measured in Utrecht, The Netherlands, and 27.1% using solar spectra and temperatures measured in Denver, Colorado.

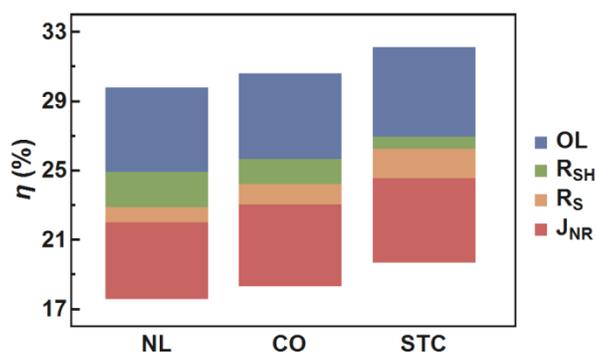

**Figure S15.** Effect of non-radiative recombination ($J_{NR}$), series resistance ($R_S$), shunt resistance ($R_{SH}$), and optical losses (OL) on the intensity-weighted power conversion efficiency over a year for the modelled perovskite solar cell, calculated using standard test conditions (STC) and solar spectra and temperatures measured in Utrecht, The Netherlands (NL)[14] and in Denver, Colorado (CO)[34].

## S6 EXPERIMENTAL

Measurements of the temperature, irradiance and solar spectrum were performed at the Utrecht Photovoltaic Outdoor Test facility (UPOT) in Utrecht, The Netherlands[14] and at the NREL Outdoor Test Facility in Denver, Colorado (US)[34] using identical MS-700 (EKO Instruments) spectroradiometers operated at 25 ± 5 °C. Light outside the spectral range of the spectroradiometers was included in the calculations by fitting measured spectra as described in [32]. In Utrecht, The Netherlands the backside-temperature of a monocrystalline Si solar panel was used to account for temperature changes of the simulated tandem solar cells. To account for temperature changes in Denver, Colorado (US) we assume that the temperature of the subcells depends on the ambient temperature $T_A$ and the solar irradiance $G$ as

$$T_C = T_A + k\,G$$

where $k$ is the Ross coefficient which is assumed to be 0.02 K m²/W.[35] Data from Denver, Colorado, US between July 8th and July 12th, 2015 are missing.